\documentclass[final]{eps}

\usepackage{graphicx}
\usepackage{times}
\usepackage{amssymb}

\volumenumber{00}
\publishyear{2012}
\frompage{\pageref{firstpage}}
\topage{\pageref{finalpage}}

\shorttitle{H. Hirashita and T. Nozawa:
Grain size distribution.}

\title{Synthesized grain size distribution in the interstellar
medium}
\author{Hiroyuki Hirashita$^1$ and Takaya Nozawa$^2$}
\affiliation{$^1$Institute of Astronomy and Astrophysics,
Academia Sinica, P.O. Box 23-141, Taipei 10617, Taiwan\\
             $^2$Kavli Institute for the Physics and Mathematics of
the Universe, Todai Institutes for Advanced Study, the University
of Tokyo, Kashiwa,\\ Chiba 277-8583, Japan}
\received{November 7, 2011}\revised{March 15, 2012}\accepted{March 16, 2012}
\abstract{
We examine a synthetic way of constructing the grain
size distribution in the interstellar medium (ISM).
First we formulate a synthetic grain size distribution
composed of three grain size distributions processed with
the following mechanisms that govern the grain size
distribution in the Milky Way: (i) grain growth by
accretion and coagulation in dense clouds,
(ii) supernova shock destruction by sputtering in
diffuse ISM, and (iii) shattering driven by turbulence in
diffuse ISM. Then, we examine if the observational grain
size distribution in the Milky Way (called MRN) is
successfully synthesized or not. We find that the three
components actually synthesize the MRN grain size
distribution in
the sense that the deficiency of small grains by (i) and
(ii) is
compensated by the production of small grains by (iii).
The fraction of each {contribution} to the total grain
processing of (i), (ii), and (iii) (i.e.,
the relative importance of the three {contributions}
to all grain processing mechanisms) is
30--50\%, 20--40\%, and 10--40\%, respectively.
We also show that the Milky Way
extinction curve is reproduced with the synthetic
grain size distributions.}

\begin{document}
\label{firstpage}
\maketitle
\copyrighttext{}

\section{Introduction}

Dust grains are important in some physical processes
in the interstellar medium (ISM). For example, they
dominate the absorption and scattering of the stellar
light, affecting the radiative transfer in the ISM.
The extinction (absorption + scattering) by dust in
the ISM as a function
of wavelength is called extinction curve
(Wickramasinghe, 1967; Hoyle and Wickramasinghe, 1991;
Draine, 2003 for review). Extinction
curves are important not only in basic radiative processes
in the ISM but also in interpreting observational data:
part of stellar light in a galaxy is scattered or absorbed
by dust grains within the galaxy in a wavelength-dependent
way according to the
extinction curve. Therefore, to derive the intrinsic stellar
spectral energy distribution of a galaxy, we always have
to correct
for dust extinction by considering the extinction curve
(Calzetti, 2001).

Extinction curves generally reflect the grain
composition and the grain size distribution.
Mathis \textit{et al.}\ (1977, hereafter MRN) show that
a mixture of silicate and graphite dust,
{as originally proposed by
Hoyle and Wickramasinghe (1969)}, with a grain size
distribution (number of grains per grain radius)
proportional to $a^{-3.5}$, where $a$ is the grain radius
($a\sim 0.001$--0.25 $\mu$m), reproduces the
Milky Way extinction curve. Pei (1992) shows that the
extinction curves in the Magellanic Clouds are also
explained by the same
power-law grain size distribution (i.e., $\propto a^{-3.5}$)
with different
abundance ratios between silicate and graphite.
Kim \textit{et al.}\ (1994) and
Weingartner and Draine (2001) have applied more
detailed fit to the Milky Way extinction curve in order
to obtain the grain size distribution. Although their
grain size distributions deviate from the MRN size
distribution, the overall trend from small to large
grain sizes roughly follows a power law
with an index near to $-3.5$. Therefore,
the MRN grain size distribution is still valid as a first
approximation of the interstellar grain size
distribution in the Milky Way.

What regulates or determines the grain size
distribution? There are some possible processes
that actively and rapidly modify the grain size
distribution. Hellyer (1970) shows that the
collisional fragmentation of dust grains finally
leads to a power-law grain size distribution
similar to the MRN size distribution (see also
Bishop and Searle, 1983). In fact,
Hirashita and Yan (2009) show that such a
fragmentation and disruption process (or shattering)
can be driven efficiently by turbulence in the
diffuse ISM.
However, they also show that grain velocities are
strongly dependent on grain size; as a result, the
grain size distribution does not converge to a
simple power-law after shattering. Moreover,
dust grains are also processed by other
mechanisms. Various authors show that the
increase of dust mass
in the Milky Way ISM is mainly governed
{by grain
growth} through the accretion of
gas phase metals onto the grains (we call
elements composing dust grains ``metals'').
(Dwek, 1998; Inoue, 2003; Zhukovska \textit{et al.},
2008; Draine, 2009; Inoue, 2011;
Asano \textit{et al.}, 2012). In
{the dense ISM}, coagulation
also occurs, making the grain sizes larger
(e.g., Hirashita and Yan, 2009). In {the} diffuse
ISM phase,
interstellar shocks associated with supernova (SN)
remnants (simply called SN shocks in this
paper) destroy dust grains,
especially small ones, by sputtering (e.g., McKee, 1989).
Shattering also occurs in SN shocks
(Jones \textit{et al.}, 1996).

Modeling the evolution of the grain size distribution
in the ISM is a challenging problem because a
variety of processes are concerned as mentioned
above. Those processes are also related to the
multi-phase nature of the ISM.
Liffman and Clayton (1989) calculate the evolution
of grain size distributions by taking into account
grain growth and shock destruction. However, their
method could not treat disruptive and coagulative
processes (i.e., shattering and coagulation).
O'Donnell and Mathis (1997) also model the
evolution of grain size distribution in a multi-phase
ISM, taking into account shattering and coagulation
in addition to the processes considered in
Liffman and Clayton (1989). They use the extinction
curve and the depletion of
gas-phase metals as quantities to be compared with
observations. Although their models are broadly
successful, the fit to the ultraviolet extinction
curve is poor, which they attribute to the
errors caused by their adopted optical constants.
They also show that
inclusion of molecular clouds in addition to
diffuse ISM phases improves the
fit to the observed depletion, but they
did not explicitly show the effects of molecular
clouds on the extinction curve.
Yamasawa \textit{et al}.\ (2011)
have recently
calculated the evolution of the grain size distribution
in the early stage of galaxy evolution
by considering the ejection of dust from SNe and
subsequent destruction in SN shocks. Since
they focus on the early stage,
they did not include other processes such as
grain growth and disruption (shattering), which
are important
in solar-metallicity environments such as in
the Milky Way (Hirashita and Yan, 2009).

Comparing theoretical grain size distributions
with observations is not a trivial procedure. In a
line of sight, we always observe a mixture of
grains processed in various ISM phases.
Therefore, a ``synthetic'' grain size distribution,
which is made by summing typical grain size
distributions in individual ISM phases with
certain weights, is to be compared with
observations. In this paper, we first formulate a
synthetic way of reproducing the grain size
distribution. Then, we  carry out a fitting of
synthetic grain size distributions to the observational
grain size distribution, in order to obtain the
relative importance of individual grain
processing mechanisms.
We do not model the multi-phase ISM in {detail},
but our {fitting contains} the information on the
weights (i.e., relative importance) of different
grain processing mechanisms, which depend on
the ISM phase.

This paper is organized as follows.
In Section \ref{sec:model}, we explain our
synthetic method to reconstruct the grain size
distribution in the ISM.
In Section \ref{sec:result}, we fit our synthetic
models to the observational grain size
distribution in the Milky Way and
examine if the fitting is successful or
not. In Section \ref{sec:discussion}, after
we discuss our results, we calculate the
extinction curves to examine if our
synthesized grain size distributions are
consistent with the observed extinction
or not. In Section \ref{sec:conclusion},
we give our conclusions.

\section{Synthetic grain size distribution}
\label{sec:model}

As explained in Introduction, we ``synthesize''
the observational grain size distribution (here,
the MRN size distribution) by summing some
representative grain size distributions in
various ISM phases. These representative
grain size distributions are explained
in Section \ref{subsec:process}.


First, the ISM is divided into two parts: one is the
part where the grain processing is occurring
(called ``grain-processing region''), and
the other is the area {where the grains already
processed in the various grain-processing regions
are well mixed
(called ``mixing region'')}.
The mass fractions of the former and the latter
regions are, respectively, $f_\mathrm{proc}$ and
$1-f_\mathrm{proc}$.
{It is reasonable to assume that the grain
size distribution in the mixing
region should be the mean grain size distribution
in the ISM (Section \ref{subsec:synthesize}).}

We assume that all grains are spherical with
material density $s$; thus, the grain mass $m$
is expressed as $m=\frac{4}{3}\pi a^3s$.
Although coagulation may produce porous
grains (e.g., Ormel \textit{et al.}, 2009), we
neglect the effects of porosity and assume
all grains to be compact.
Two grain species are treated in this paper;
silicate and graphite.
{To avoid complexity arising from compound species,
we treat these two species separately.
This separate treatment is also practical in
this paper as we (and other authors usually)
assume that the observed extinction curve
can be fitted with the two species
(Section \ref{subsec:extinction}).}
Before being processed, the grain size
distribution is assumed to be MRN:
a power-law function with
power index $-r$ ($r=3.5$), and
upper and lower bounds for the grain radii
(whose values are determined below)
$a_\mathrm{min}$ and $a_\mathrm{max}$, respectively:
\begin{eqnarray}
n_\mathrm{MRN}(a)=
{\displaystyle
\frac{(4-r)\rho_\mathrm{d}}
{\frac{4}{3}\pi s(a_\mathrm{max}^{4-r}-
a_\mathrm{min}^{4-r})n_\mathrm{H}}
}\, a^{-r}
\end{eqnarray}
for $a_\mathrm{min}\leq{a}\leq a_\mathrm{max}$.
If $a<a_\mathrm{min}$ or $a>a_\mathrm{max}$,
$n_\mathrm{MRN}(a)=0$. The grain size
distribution is defined so that
$n_\mathrm{MRN}(a)\, da$ is the number of
grains whose sizes are between $a$ and
$a+da$ per hydrogen nucleus. The dust mass
density, $\rho_\mathrm{d}$,
is related to the metallicity $Z$
{(the mass fraction of elements heavier than
helium in the ISM)} and
the hydrogen number density $n_\mathrm{H}$ as
(Hirashita and Kuo, 2011)
\begin{eqnarray}
\rho_\mathrm{d}=
\frac{m_\mathrm{X}}{f_\mathrm{X}}
(1-\xi )\left(\frac{Z}{\mathrm{Z}_{\odot}}\right)
\left(\frac{\mathrm{X}}{\mathrm{H}}\right)_{\odot}
n_\mathrm{H},\label{eq:rho_d}
\end{eqnarray}
where $m_\mathrm{X}$ is the atomic mass of the
key element X (X = Si for silicate and C for graphite),
$f_\mathrm{X}$ is the mass fraction of X in the
dust, {$\xi$  is the fraction of element X in
gas phase (i.e., the fraction $1-\xi$ is in dust
phase)}, and (X/H)$_{\odot}$ is
the solar abundance relative to hydrogen in
number density.
The metallicity is assumed to be solar
($Z=\mathrm{Z}_\odot$).

We fix the maximum grain radius as
$a_\mathrm{max}=0.25~\mu$m (MRN).
Although the lower bound of the grain size is
poorly determined from the extinction curve
(Weingartner and Draine, 2001), we assume
that $a_\mathrm{min}=0.3$ nm, since
a large number of very small grains are indeed
necessary to explain the mid-infrared excess of
the dust emission in the Milky Way
(Draine and Li, 2001).
For the other parameters, we follow
Hirashita (2012). We assume
that 0.75 of Si is condensed into silicate
(i.e., $\xi =0.25$)
while 0.85 (i.e., $\xi =0.15$) of C is
included into graphite. Those values are
roughly consistent with the observed depletion
(e.g., Savage and Sembach, 1996), and
reproduce the Milky Way extinction curve
(Section \ref{subsec:extinction}). We adopt
the following abundances for Si and C:
$(\mathrm{Si/H})_\odot=3.55\times 10^{-5}$
and
$(\mathrm{C/H})_\odot=3.63\times 10^{-4}$.
We assume that Si occupies a mass
fraction of 0.166 ($f_\mathrm{X}=0.166$)
in silicate while
C is the only element composing
graphite ($f_\mathrm{X}=1$).
We adopt $s=3.3$ and 2.26~g~cm$^{-3}$
for silicate and graphite, respectively.

By using the MRN size distribution as the
initial condition, we calculate the
evolution of grain size distribution by
the various processes treated in
Section \ref{subsec:process}.
In the numerical calculation,
the grains going out of
the radius range between $a_\mathrm{min}$
and $a_\mathrm{max}$ are removed from
the calculation ({the removed mass
fraction is $<1$\%}).
The processes considered
are (i) ``grain growth'' -- grain
growth by accretion and coagulation in
dense medium,
(ii) ``shock destruction'' -- destruction by
sputtering in SN shocks, and
(iii) ``grain disruption'' -- grain disruption
by shattering in interstellar
turbulence. Shattering in
SN shocks (Jones \textit{et al.}, 1996)
could be included as a separate
component, {but 
in our framework, it is not possible
to separately constrain the contributions
from the two shattering mechanisms because
both shattering mechanisms
(turbulence and SN shocks) selectively
destroy grains with $a\gtrsim 0.03~\mu$m
and increase smaller grains, predicting
similar grain size distributions. Thus,}
we simply assume that the size distribution
of shattered grains, whatever the shattering
mechanism may be, is represented by the
one adopted in Section \ref{subsubsec:shattering}.

Although our fitting procedures are based on
grain size distributions, we should keep in
mind that observational constraints on the
grain size distribution is mainly obtained by
extinction curves. Weingartner \& Draine (2001)
performed a detailed fit to the Milky Way
extinction curve. However, the grain size
distributions derived by them broadly follow
an MRN-like power law, although there are
bumps and dips at some sizes.
We will examine the consistency with the
extinction curve later in
Section \ref{subsec:extinction}.

\subsection{Processes considered}
\label{subsec:process}

\subsubsection{Grain growth}
\label{subsubsec:growth}

Grain growth occurs in the dense ISM, especially
in molecular clouds, through the accretion of
metals (called accretion) and the sticking of
grains (called coagulation). The change of
grain size distribution by grain
growth has been considered in
our previous paper (Hirashita, 2012).
The grain size distribution after grain
growth is denoted as
$n_\mathrm{grow}(a,\, t_\mathrm{grow})$,
where $t_\mathrm{grow}$ is the duration
of grain growth. We adopt
$t_\mathrm{grow}=10$ and 30 Myr based
on typical lifetime of molecular clouds
(e.g., Lada \textit{et al.}, 2010).
The metallicity in the Milky Way is high
enough to allow complete depletion of
grain-composing materials onto dust
grains in $\sim 10$ Myr. Thus, the total masses of
silicate and graphite become 1.33
($=1/0.75$) and 1.18 ($=1/0.85$) times
as large as the initial values, respectively
{(recall that the dust mass becomes
$1/(1-\xi )$ times as much if all the
dust grains accrete all the gas-phase metals)}.
The difference in the grain size distribution
between $t_\mathrm{grow}=10$ and 30 Myr is
predominantly caused by coagulation
rather than accretion.

\subsubsection{Shock destruction}
\label{subsubsec:shock}

We calculate the change of grain size distribution
by SN shock destruction in a medium swept by
a SN shock, following
Nozawa \textit{et al}.\ (2006).
{All SN explosions are represented by
an explosion of a star which has a
mass of  20 M$_\odot$ at the zero-age
main sequence, and the SN explosion
energy is assumed to be $10^{51}$ erg. For the
ISM, we adopt a hydrogen number density of
0.3 cm$^{-3}$
(since the destruction is predominant in
the diffuse ISM; McKee, 1989), and solar
metallicity.
The calculation of grain destruction
is performed until the shock velocity is
decelerated down to 100 km s$^{-1}$
($8\times 10^4$ yr after the explosion).}
We apply the material properties of
Mg$_2$SiO$_4$ and carbonaceous dust
in Nozawa \textit{et al.}\ (2006) for
silicate and graphite, respectively.
We denote
the grain size distribution after shock
destruction by $n_\mathrm{shock}(a)$.
The destroyed mass fractions of silicate and
graphite are 0.38 and 0.27, respectively.


\subsubsection{Disruption}
\label{subsubsec:shattering}

Grain motions driven by interstellar turbulence
lead to grain disruption (shattering) in the
diffuse ISM (Yan et al., 2004;
Hirashita and Yan, 2009). Among the various ISM
phases, dust grains can
acquire the largest velocity dispersion in
a warm ionized medium (WIM). We recalculated
the results of earlier workers based on our
assumed initial conditions.
We adopt the same grain velocity dispersions
and hydrogen number density
($n_\mathrm{H}=0.1$ cm$^{-3}$) in
the WIM as adopted in Hirashita and Yan (2009).
{The fragments are assumed to follow a
power-law size distribution with a power
index of $-3.3$ (Jones \textit{et al.}, 1996;
Hirashita and Yan, 2009).}
We denote the grain size distribution after
disruption as
$n_\mathrm{disr}(a,\, t_\mathrm{disr})$,
where $t_\mathrm{disr}$ is the duration of
shattering in the WIM.
The lifetime of WIM is
estimated to be a few Myr from the
recombination timescale and the lifetime
of ionizing stars (Hirashita and Yan, 2009).
Thus, we adopt
$t_\mathrm{disr}=3$ and 10 Myr for our
calculation in causing
moderate and significant disruption.

\subsection{Synthesizing the grain size
distribution}\label{subsec:synthesize}

In the beginning of this section, we introduced
the mass fraction ($f_\mathrm{proc}$) of ISM
hosting grains which are now being processed
(``grain-processing region''). The mean grain
size distribution over all the
grain-processing region, $n_\mathrm{synt}(a)$,
can be synthesized with the processed grain
size distributions,
{
$n_\mathrm{grow}(a,\, t_\mathrm{grow})$
(grain size distribution after grain growth
with a growth duration of $t_\mathrm{grow}$),
$n_\mathrm{shock}(a)$
(grain size distribution after shock
destruction), and
$n_\mathrm{disr}(a,\, t_\mathrm{disr})$
(grain size distribution after disruption
with a shattering duration of $t_\mathrm{disr}$)}:
\begin{eqnarray}
n_\mathrm{synt}(a) & = & f_\mathrm{grow}
n_\mathrm{grow}(a,\, t_\mathrm{grow})+
f_\mathrm{disr}
n_\mathrm{disr}(a,\, t_\mathrm{disr})\nonumber\\
& & + f_\mathrm{shock}n_\mathrm{shock}(a),
\label{eq:sum_proc}
\end{eqnarray}
where $f_\mathrm{grow}$, $f_\mathrm{disr}$
and $f_\mathrm{shock}$ are the mass fractions
of medium hosting, respectively, grain growth,
disruption, and grain destruction in the
grain-processing region. We call
$n_\mathrm{synt}(a)$ ``synthetic grain size
distribution''.
{If both species are spatially well mixed,
they would have common values for $f_\mathrm{grow}$,
$f_\mathrm{shock}$, and $f_\mathrm{disr}$.}

The mean grain size distribution in the ISM
is denoted as $n_\mathrm{mean}(a)$ and
expressed as
\begin{eqnarray}
n_\mathrm{mean}(a)=(1-f_\mathrm{proc})
n_\mathrm{mean}(a)+f_\mathrm{proc}
n_\mathrm{synt}(a),
\end{eqnarray}
since it is assumed that the grain size
distribution in the mixing region has already
become the mean grain size distribution.
{By assumption},
the mean size distribution is
MRN: $n_\mathrm{mean}(a)=n_\mathrm{MRN}(a)$.
This condition is equivalent
to
\begin{eqnarray}
n_\mathrm{synt}(a)=n_\mathrm{MRN}(a).
\end{eqnarray}

In the Milky Way ISM, since the grain mass is
roughly in equilibrium between the growth in
clouds and the destruction by SN shocks
(Inoue, 2011),
we apply
$f_\mathrm{grow}R_1=f_\mathrm{shock}
R_2$,
where $R_1$ is the fraction of dust mass growth
in clouds (0.33 and 0.18 for silicate and
graphite, respectively;
Section \ref{subsubsec:growth}),
and $R_2$ is the destroyed fraction of dust
in a SN blast (0.38 and 0.27 for silicate and
graphite, respectively; Section \ref{subsubsec:shock}). Thus, we put a
constraint,
\begin{eqnarray}
f_\mathrm{shock}=(R_1/R_2)
f_\mathrm{grow}.\label{eq:equil}
\end{eqnarray}
We approximately adopt $R_1/R_2=0.8$
as a mean value between silicate and graphite.
{As mentioned above, if the two species
(silicate and graphite) are spatially well
mixed, both species would have common
values for
$f_\mathrm{grow}$, $f_\mathrm{shock}$, and
$f_\mathrm{disr}$.
Thus, we adopt 
a single value for $R_1/R_2$.}

Using the above constraints, Eq.~(\ref{eq:sum_proc})
is reduced to
\begin{eqnarray}
n_\mathrm{synt}(a)=f_\mathrm{grow}
n_\mathrm{g,s}(a)+f_\mathrm{disr}
n_\mathrm{disr}(a,\, t_\mathrm{disr}),
\end{eqnarray}
where
$n_\mathrm{g,s}(a)\equiv
n_\mathrm{grow}(a,\, t_\mathrm{grow})
+(R_1/R_2)n_\mathrm{shock}(a)$.
Thus, we treat $f_\mathrm{grow}$ and
$f_\mathrm{disr}$ as free parameters.
{We define
the sum of all the fractions as
\begin{eqnarray}
f_\mathrm{tot} & \equiv & f_\mathrm{grow}+
f_\mathrm{shock}+f_\mathrm{disr}\nonumber\\
& = &
\left(1+\frac{R_1}{R_2}\right)f_\mathrm{grow}
+f_\mathrm{disr}.\label{eq:constraint}
\end{eqnarray}
If the grain size distribution
is predominantly modified by  the three processes
considered in this paper, we expect
that $f_\mathrm{tot}=1$. The deviation of
$f_\mathrm{tot}$ from 1 is an indicator of
goodness of our assumption that the grain
size distribution is modified by the three
processes.
}

\subsection{Best fitting parameters}

We search for a set of parameters,
$(f_\mathrm{grow}, f_\mathrm{disr})$,
which minimizes
the square of the difference:
\begin{eqnarray}
\delta^2=\frac{1}{N}\sum_{i=1}^N\, [\log n_\mathrm{synt}(a_i) -\log
n_\mathrm{MRN}(a_i)]^2,
\end{eqnarray}
where $a_i$ is the grain size sampled by logarithmic
bins (i.e., $\log a_{i+1}-\log a_i$ is the same for any $i$),
and $N$ is the number of the sampled grain radii
($N=512$ in our model, but the results are insensitive to
$N$).

\begin{figure*}[t]
\includegraphics[width=0.48\textwidth]{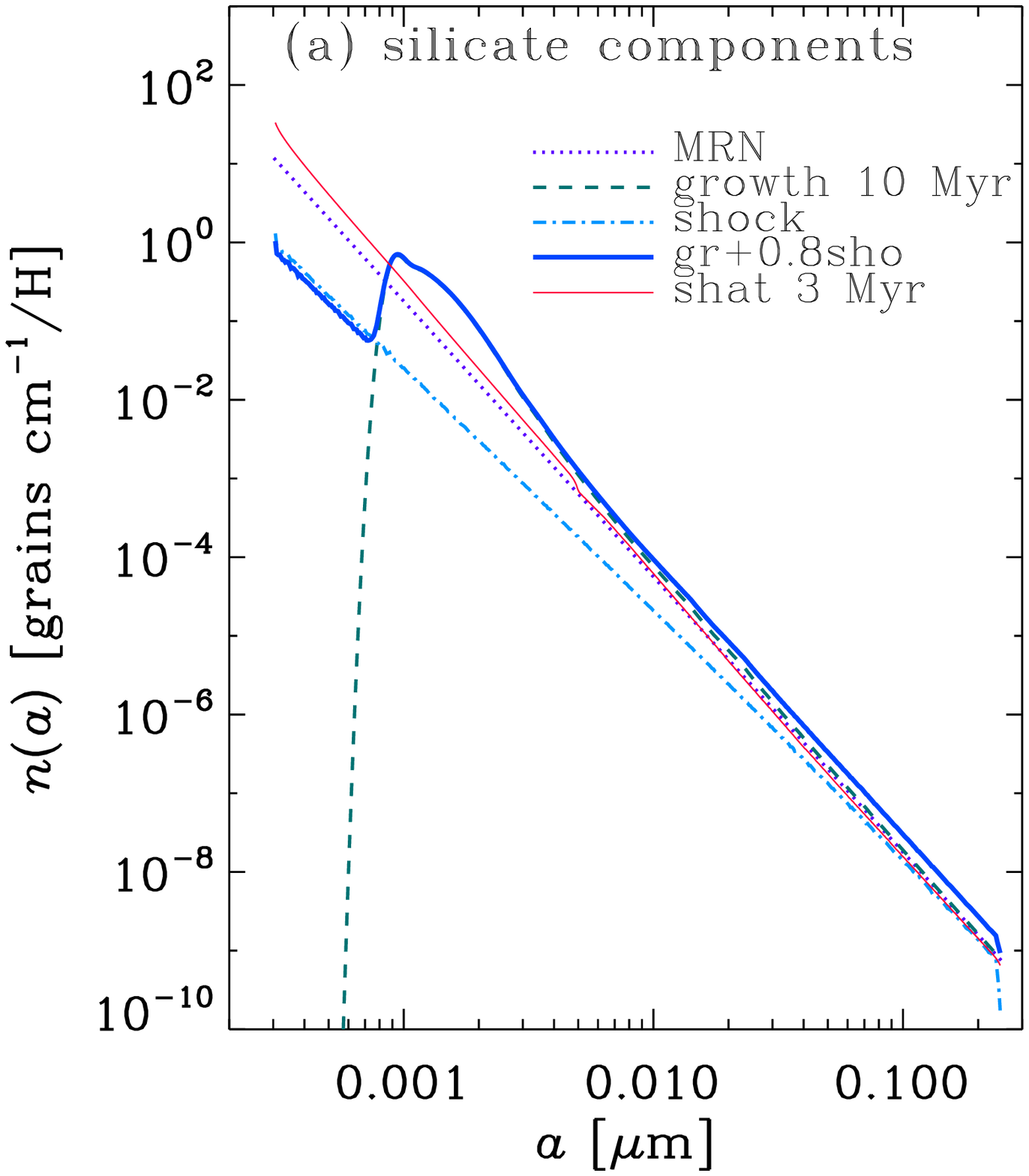}
\includegraphics[width=0.48\textwidth]{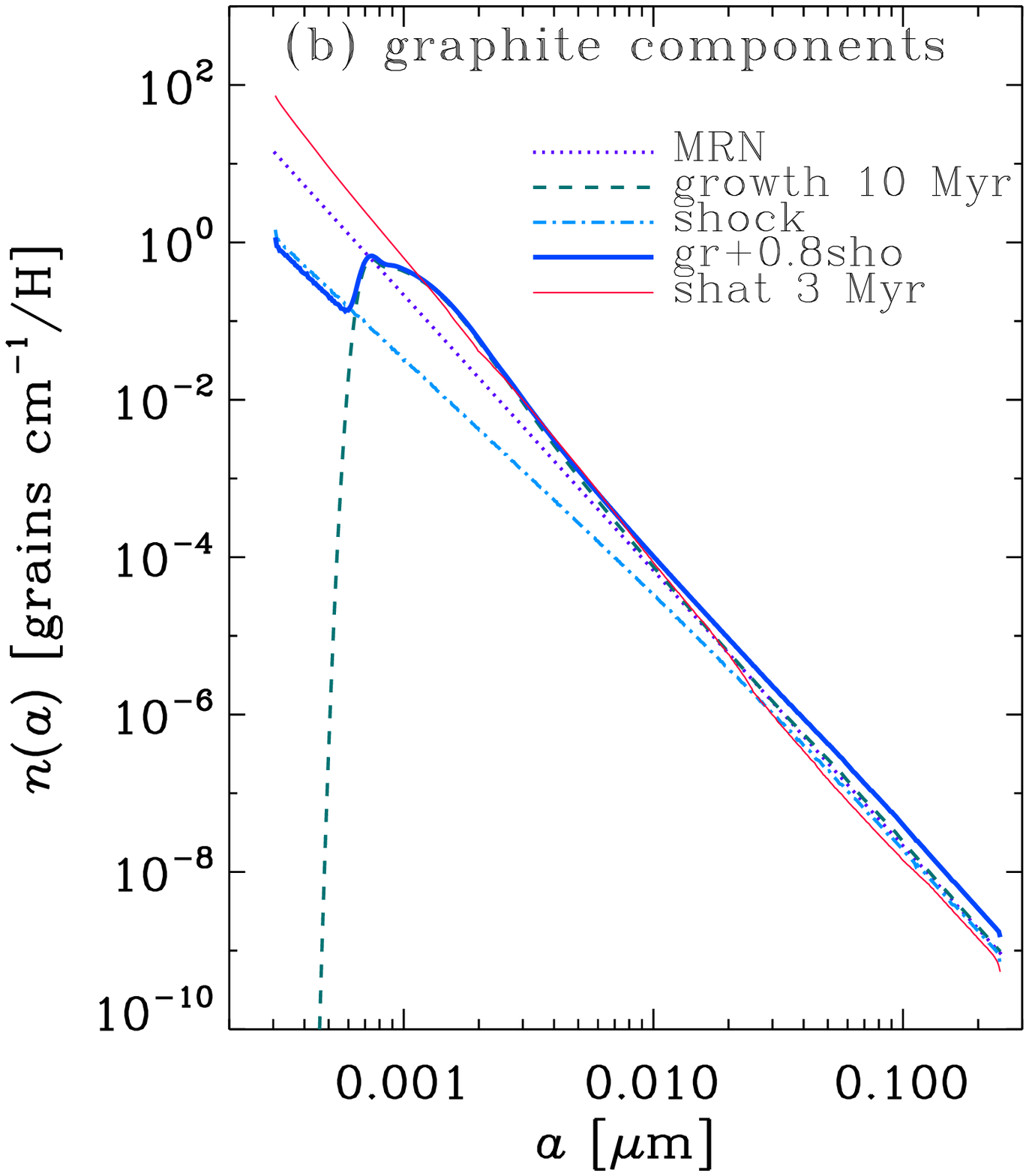}\\
\includegraphics[width=0.48\textwidth]{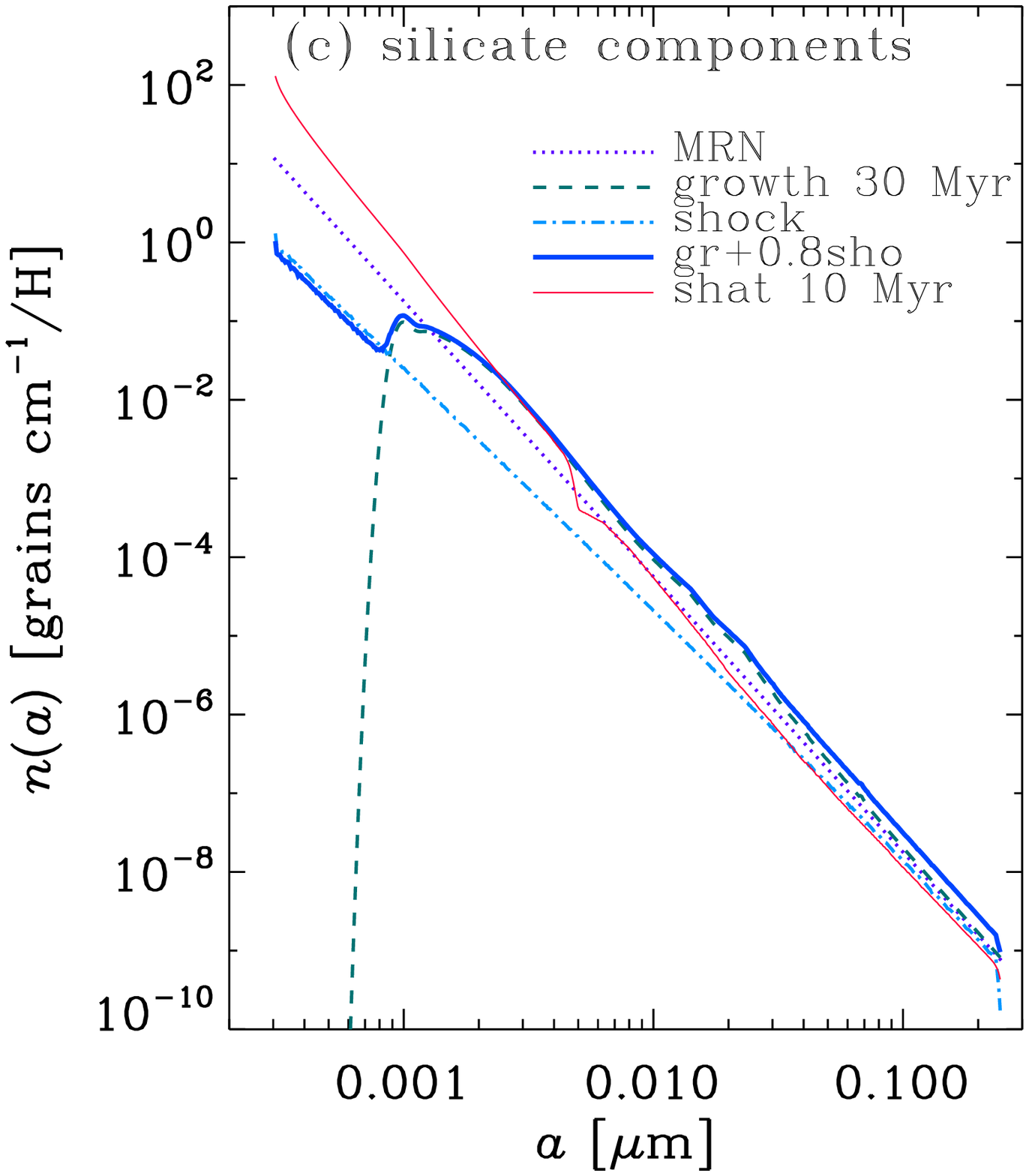}
\includegraphics[width=0.48\textwidth]{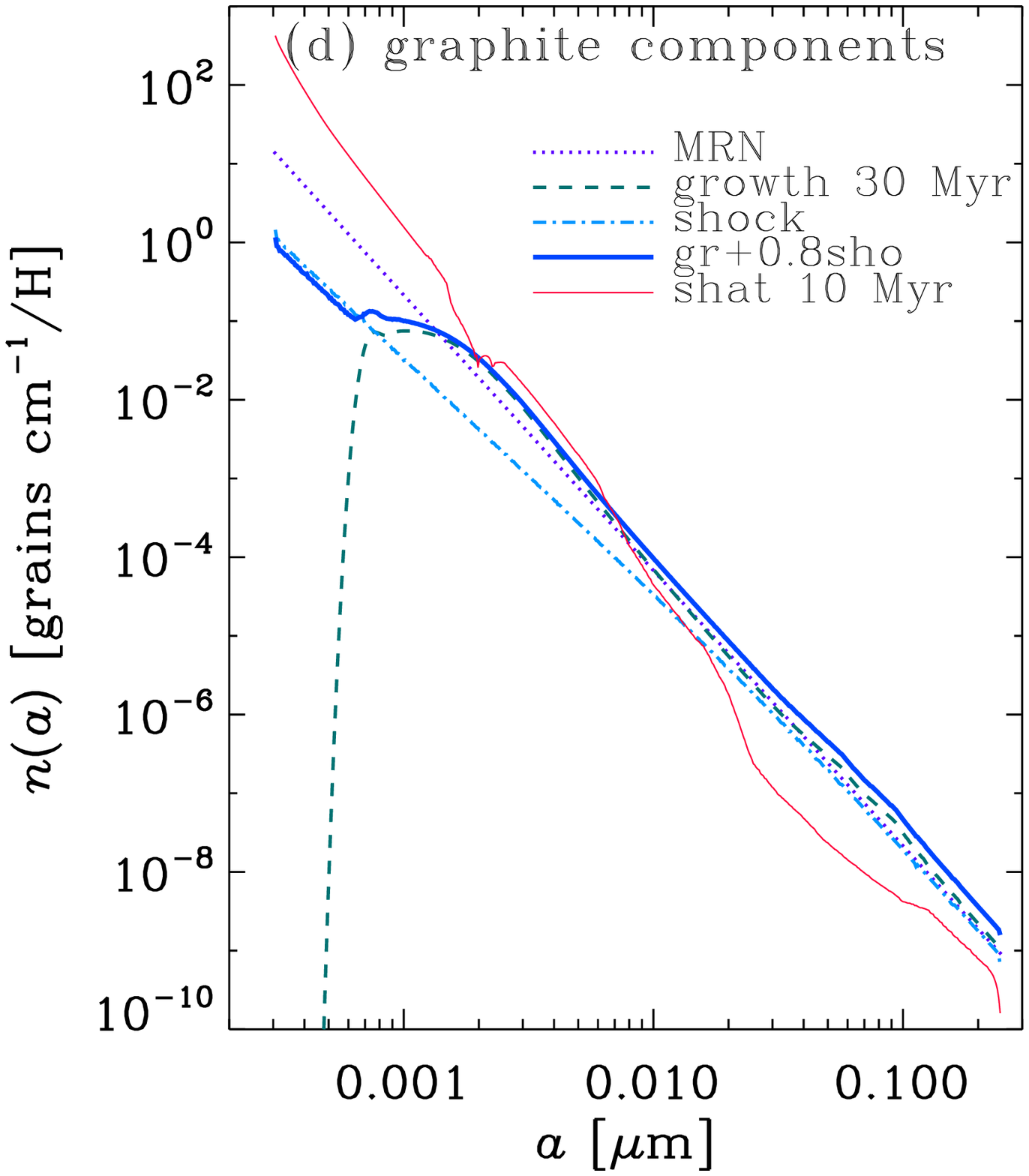}
\caption{\footnotesize
Individual components for synthesized grain size
distributions. The thin solid, dashed, and
dot-dashed lines represent individual components
processed by disruption (shattering) for 3 Myr in
Panels (a) and (b),
and for 10 Myr in Panels (c) and (d), growth for
10 Myr in Panels (a) and (b) and for 30 Myr in
Panels (c) and (d), and shock, respectively.
The thick solid line shows
$n_\mathrm{g,s}(a)=n_\mathrm{grow}(a,\, t_\mathrm{grow})+
0.8n_\mathrm{shock}(a)$. The dotted line shows the
MRN size distribution adopted in this paper.
Panels (a) and (c) present silicate while
Panels (b) and (d) show graphite.
}
\label{fig:component}
\end{figure*}

The individual components of processed grain
size distributions
[$n_\mathrm{grow}(a,\, t_\mathrm{grow})$,
$n_\mathrm{shock}(a)$, and
$n_\mathrm{disr}(a,\, t_\mathrm{disr})$]
as well as $n_\mathrm{g,s}(a)$
for silicate and
graphite are shown in Fig.\ \ref{fig:component}.
The MRN size distribution, which should be fitted,
is also presented.
{Our fitting procedure is first applied
separately for silicate and graphite, although
we discuss a possibility that both species have
common values for
$(f_\mathrm{grow}, f_\mathrm{disr})$
later in Section \ref{sec:discussion}.}

\section{Results}\label{sec:result}

In Table \ref{tab:model}, we show the best-fitting values
of $f_\mathrm{grow}$ and $f_\mathrm{disr}$. We examine
$t_\mathrm{grow}=30$ and 10 Myr, and
$t_\mathrm{disr}=3$ and 10 Myr as mentioned in
Section \ref{subsec:process}. We observe that
$f_\mathrm{grow}=0.16$--0.55
and $f_\mathrm{disr}=0.06$--0.57
fit the MRN grain size distribution.
{The sum of all the fractions,
$f_\mathrm{tot}$ (Eq.\ \ref{eq:constraint})
is unity with the maximum difference of 15\%
(see the column of
$f_\mathrm{tot}$ in Table \ref{tab:model}).}

\begin{table*}[t]
\renewcommand{\arraystretch}{1.2}
\vspace{-.3cm}
\caption{Models.}
\vspace{-.1cm}
\begin{center}
\begin{tabular}{cccccccc} \hline
Name & species & $t_\mathrm{grow}$ &
$t_\mathrm{disr}$ & $f_\mathrm{grow}$ &
$f_\mathrm{disr}$ & $\delta^2$ & $f_\mathrm{tot}$\\
 & & (Myr) & (Myr) & & & ($10^{-3}$) & \\ \hline
A & silicate & 30 & 3 & 0.26 & 0.46 & 3.7 & 0.93\\
   & graphite &  & & 0.43 & 0.24 & 4.2 & 1.01 \\ \hline
B & silicate & 10 & 3 & 0.16 & 0.57 & 16 & 0.86\\
   & graphite &  &  &  0.40 & 0.20 & 13 & 0.92\\ \hline
C & silicate & 30 & 10 & 0.42 & 0.16 & 6.4 & 0.92\\
   & graphite & & & 0.55 & 0.073 & 7.3 & 1.06\\ \hline
D & silicate & 10 & 10 & 0.40 & 0.15 & 33 & 0.88\\
   & graphite &   & & 0.50 & 0.057 & 15 & 0.96\\ \hline
\end{tabular}

\vspace{.1cm}
{Note: $f_\mathrm{shock}=0.8f_\mathrm{grow}$ from
Eq.\ (\ref{eq:equil}).}
\end{center}
\label{tab:model}
\end{table*}

\begin{figure*}[t]
\includegraphics[width=0.48\textwidth]{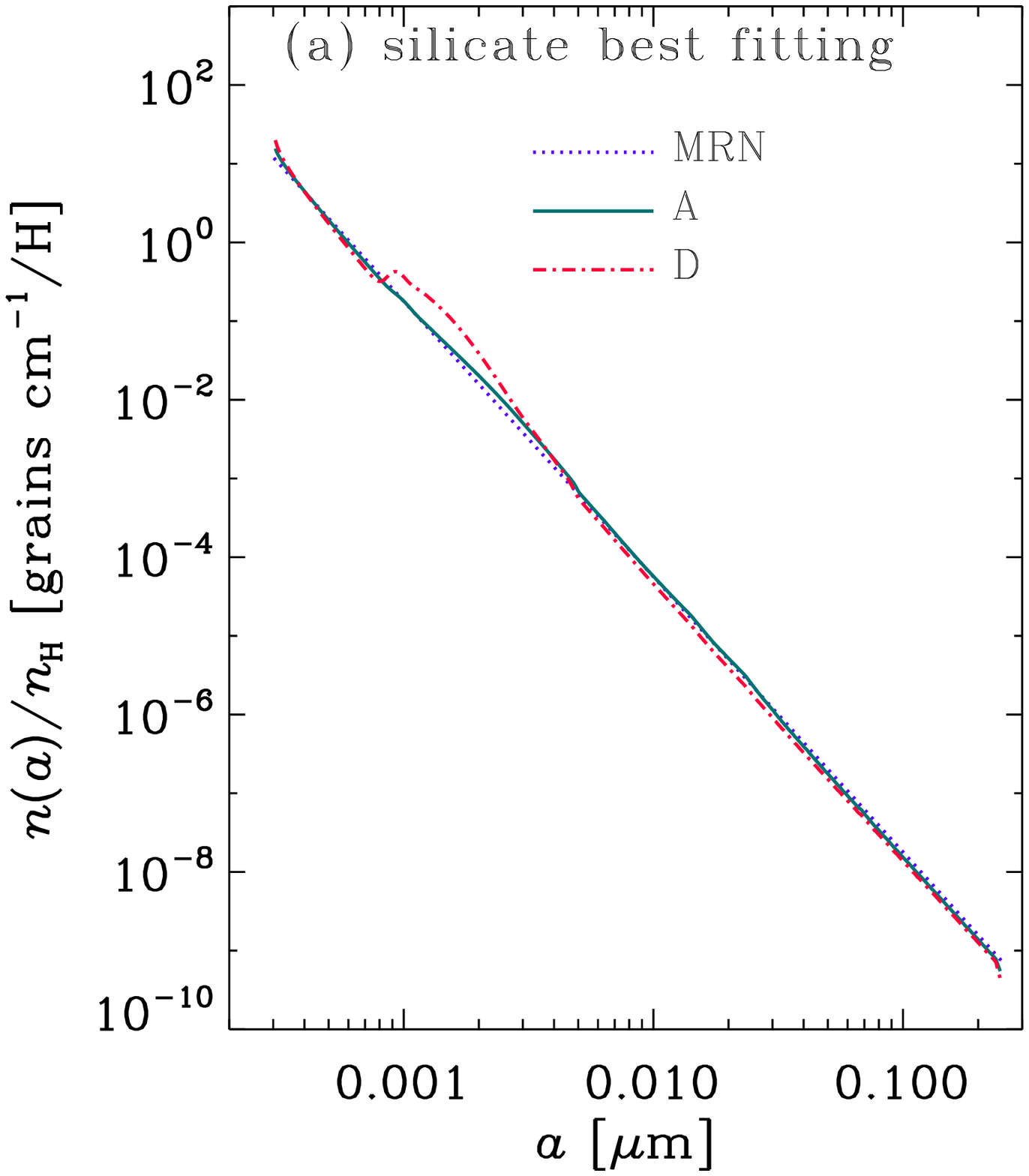}
\includegraphics[width=0.48\textwidth]{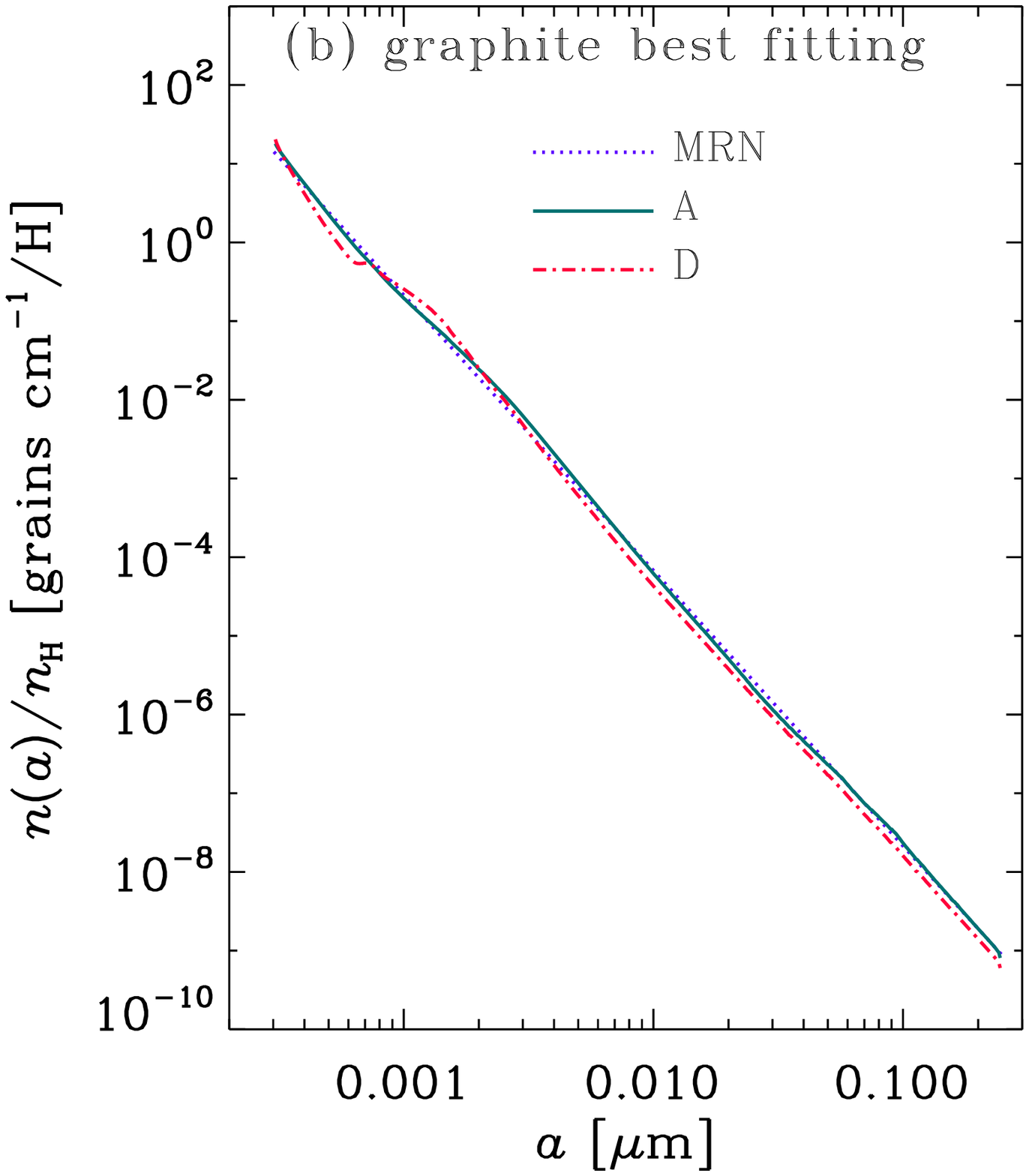}
\caption{\footnotesize
Best-fitting synthetic grain size distributions to the MRN
size distribution. Panels (a) and (b) show silicate and
graphite, respectively. We only show two models
(A and D; solid and dot-dashed lines, respectively)
for the smallest and largest residuals ($\delta^2$) among
the four models.
The dotted line shows the MRN size distribution.
}
\label{fig:synthesized}
\end{figure*}

In order to show how the synthetic grain size
distributions reproduce the MRN size distribution,
we present Fig.~\ref{fig:synthesized}, where we
only show Models A and D for the smallest and
the largest residuals $\delta^2$. We observe that
{the best-fitting
results} are fairly consistent with the
MRN size distribution. In particular, the
enhanced and depleted abundances of small grains at
$a\lesssim 0.001~\mu$m in $n_\mathrm{disr}$ and
$n_\mathrm{g,s}$,
respectively, cancel out very well, especially
in Model A. In Model D, the synthesized size
distribution slightly fails to fit the MRN around
$a\sim 0.001$--$0.002~\mu$m because both $n_\mathrm{g,s}$
and $n_\mathrm{disr}$ (which are used for the
fitting) show an excess around this grain radius range;
thus, the excess around
these sizes in the synthesized grain size
distribution inevitably remains in Model D.
However, the Milky Way extinction curve
is reproduced even by Model D within a difference of
$\sim 10$\% (see Section \ref{subsec:extinction}).

\begin{figure*}[t]
\includegraphics[width=0.48\textwidth]{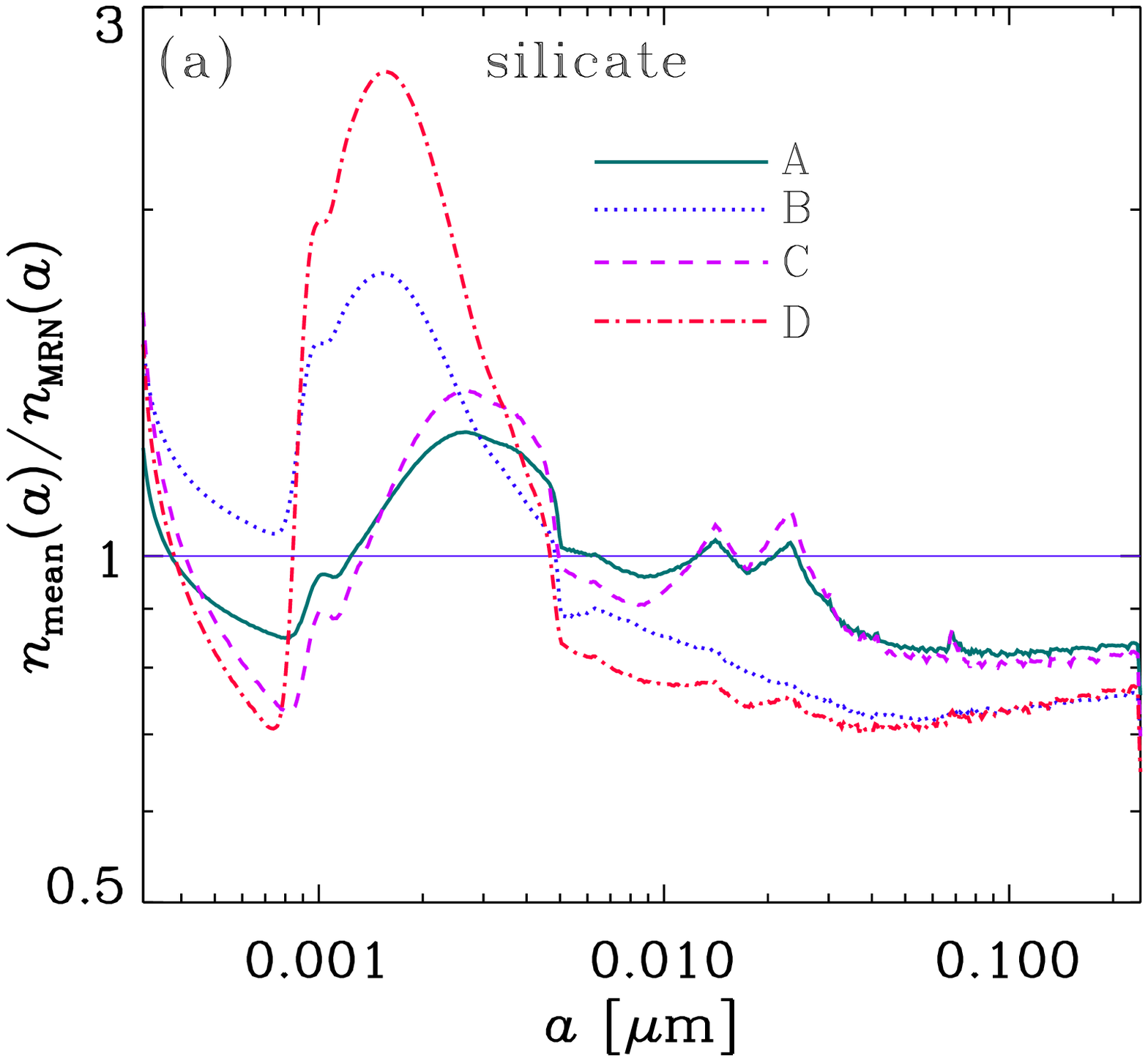}
\includegraphics[width=0.48\textwidth]{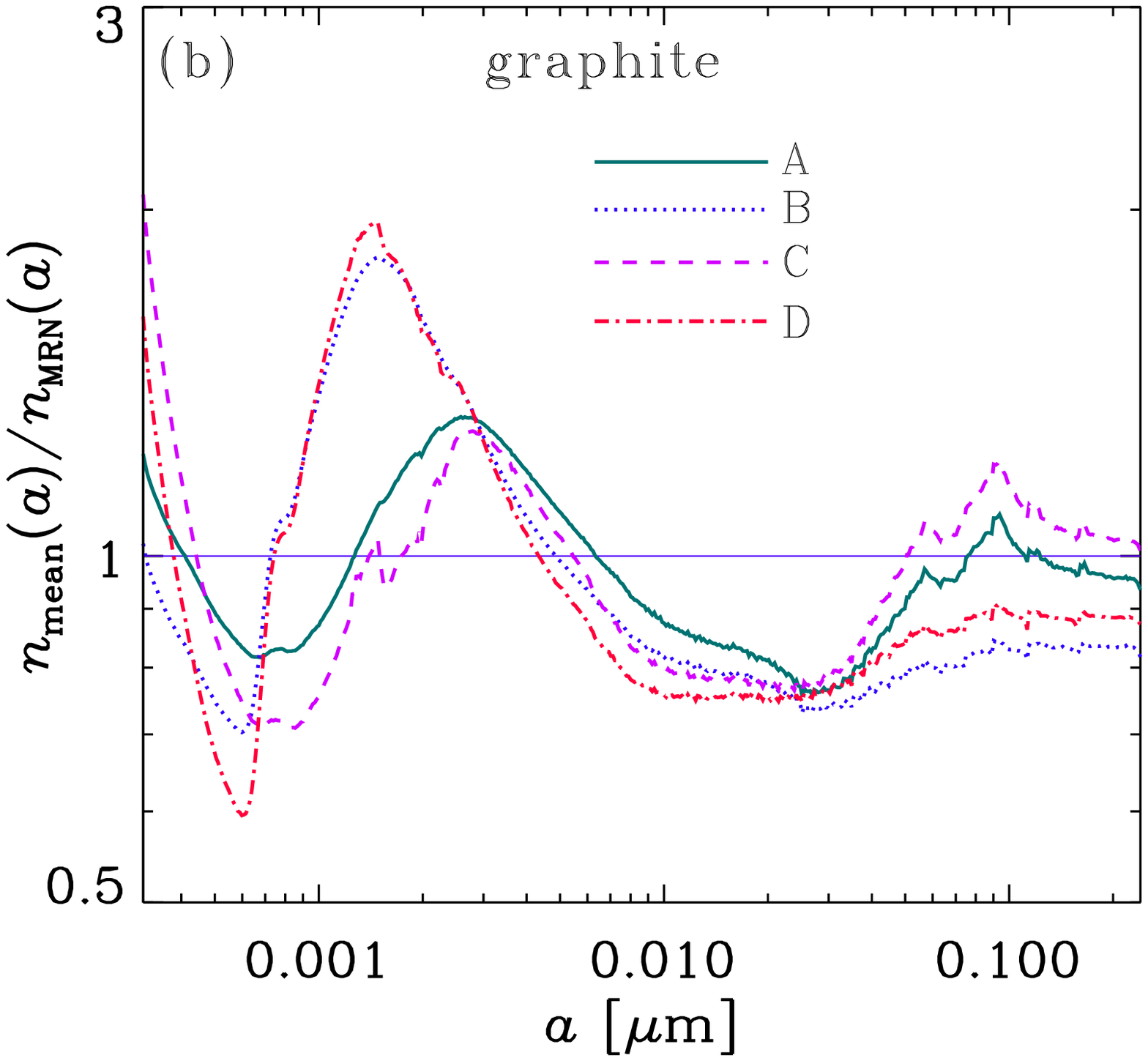}
\caption{\footnotesize
The ratio of the synthetic grain size distribution
to the MRN size distribution. Panels (a) and (b) show
silicate and
graphite, respectively. The solid, dotted, dashed, and
dot-dashed lines represent Models A, B, C, and D,
respectively.
}
\label{fig:ratio}
\end{figure*}

In order to see the details of the fitting, we show
the ratio between the synthesized grain size
distribution and the MRN distribution in
Fig.\ \ref{fig:ratio}. There is a general trend of
excess around $a\sim 0.001$--0.003 $\mu$m,
which is due to grain
growth (see Fig.\ \ref{fig:component}).
The excess is stronger in Models B and D
than in Models A and C, which is why the fit is worse in
Models B and D than Models A and C
(Table \ref{tab:model}). Since the bump comes from
grain growth, the fit tends to suppress $f_\mathrm{grow}$
in the presence of a strong bump. As a result,
$f_\mathrm{tot}$ is smaller in Models B and D than
Models A and C (Table \ref{tab:model}).
We also observe in Fig.\ \ref{fig:ratio} that
the bump appears at different grain radii between
Models A/C and B/D because of the difference in
the duration of grain growth.
{
This bump may disappear if coagulation is more
efficient than assumed here: more efficient coagulation
may be realized if
$t_\mathrm{grow}\gg 30$ Myr
and/or coagulation also occurs in denser regions
(Section \ref{subsec:stellar}).
}

Fig.\ \ref{fig:ratio} also indicates that the synthetic
grain size distributions tend to be
deficient at the largest grain sizes
($a\gtrsim 0.1~\mu$m). This is because
shattering tends to process large grains into
small sizes (see Fig.\ \ref{fig:component}).
The deficiency of large grains may be overcome
if we include the supply of large grains by stellar
sources of efficient coagulation as discussed in
Section \ref{subsec:stellar}.
Because of significant grain growth in Models A and C,
the deficiency of large grains is recovered by
grain growth at $a\sim 0.01$--$0.03~\mu$m for silicate.
In Models A and C of graphite, shattering causes a
dip feature around $a\sim 0.03~\mu$m as seen in
Fig.\ \ref{fig:component}, which also appears in
Fig.\ \ref{fig:ratio}. In the WIM,
where shattering is assumed to occur in this paper,
grains with
$a\gtrsim\mbox{a few}\times 10^{-2}~\mu$m are
accelerated up to velocities larger than
the shattering threshold by turbulence.
Shattering efficiently destroys small grains because of
their large surface-to-volume ratios. Thus, the
shattering efficiency is the largest
for the smallest grains that {attain}
a velocity above the shattering threshold.
This is the reason why the grains around
$a\sim 0.03~\mu$m are particularly destroyed by
shattering. This dip feature
would be smoothed out in reality since the
grain velocity driven by turbulence has a
dependence on grain charge, gas density, magnetic field,
etc., all of which have {a wide range} within a galaxy.

\section{Discussion}\label{sec:discussion}

\subsection{Derived parameters}

The obtained values of the parameters
$f_\mathrm{grow}$ and $f_\mathrm{disr}$
reflect the fraction of individual grain processing
mechanisms. In other words, these two quantities
show the relative importance of grain growth and
disruption. Note that the efficiency of shock
destruction is automatically constrained by the
balance with the mass growth by grain growth
(Eq.\ \ref{eq:equil}). Table \ref{tab:model} shows
that the best-fitting
parameters are not very sensitive to $t_\mathrm{grow}$
(duration of grain growth) but that they are sensitive
to $t_\mathrm{disr}$. For larger $t_\mathrm{disr}$,
only a smaller $f_\mathrm{disr}$ is necessary
because the grain size distribution is more modified.
As expected, $f_\mathrm{disr}t_\mathrm{disr}$
is less sensitive to $t_\mathrm{disr}$; note that
$f_\mathrm{disr}t_\mathrm{disr}$
is the mean duration of disruption per processed grain.

Seeing all the models, we find
$f_\mathrm{grow}\sim 0.2$--0.6 and
$f_\mathrm{disr}\sim 0.06$--0.6 (or
$f_\mathrm{disr}t_\mathrm{disr}\sim 0.6$--1.7 Myr).
As mentioned in Section \ref{subsec:synthesize}, if
silicate and graphite are well mixed in the ISM,
they are expected to have the common values for
$f_\mathrm{grow}$, $f_\mathrm{shock}$, and
$f_\mathrm{disr}$. In this sense, Models C and
D work better than Models A and B. In summary,
20--60\% of processing occurs in dense clouds (i.e.,
grain growth), while a processed dust grain
experiences disruption for $\sim 1$ Myr on average
(or disruption accounts for 6--60\% of processing). From the
equilibrium constraint of the total dust mass
(Eq.\ \ref{eq:equil}), the fraction of shock destruction
to all the processing is
$f_\mathrm{shock}=0.8f_\mathrm{grow}\sim 0.1$--0.4.

{
The sum of all the fractions, $f_\mathrm{tot}$
(Eq.\ \ref{eq:constraint}), is unity with the
maximum deviation of 15\%.
In other words, we cannot
reject other processing mechanisms, which could
contribute to the grain processing with $\lesssim 15$\%.
}

We have shown that the grain size distributions
after (i) grain growth,
(ii) shock destruction, and (iii) grain disruption
can synthesize the MRN size distribution. It is also
likely that we can say the opposite; that is,
to realize the MRN size distribution, those three
processes are crucial. Without (i) the grain mass
just decreases; without (ii) the grain mass just
increases; without (iii) there is no mechanism
that produce the large abundance of small grains.

\subsection{Stellar sources?}\label{subsec:stellar}

In this paper, we did not consider dust supply
from stars, because the dust mass in the Milky Way
is governed by the equilibrium between grain
growth in molecular clouds and grain destruction
by SN shocks (e.g.\ Inoue, 2011). Dust grains
supplied from stars may
be biased to large sizes. Production of
large grains from asymptotic giant branch
(AGB) stars is indicated observationally
(Groenewegen, 1997;
Gauger \textit{et al.}, 1998;
H\"{o}fner, 2008; Mattsson \& H\"{o}fner, 2011).
The dust ejected from SNe is also biased to
large grain sizes because small grains are selectively
destroyed by the shocked region within the SNe
(Bianchi and Schneider, 2007;
Nozawa \textit{et al.}, 2007).
Coagulation associated with
star formation is also a source of large grains if
coagulated grains in circumstellar
environments are somehow ejected into
the ISM. For this possibility, Hirashita and Omukai
(2009) have shown that dust grains can grow up to
micron sizes by coagulation in star formation
(see also Ormel \textit{et al.}, 2009).
{
As mentioned in Section \ref{sec:result},
efficient coagulation may also solve the bump problem
around $a\sim 0.001$--0.003 $\mu$m.
}
These possible sources of large grains may be worth
including in dust evolution models in the future.


\begin{table*}[t]
\renewcommand{\arraystretch}{1.2}
\vspace{-.3cm}
\caption{Models with $f_\mathrm{tot}=1$.}
\vspace{-.1cm}
\begin{center}
\begin{tabular}{cccccccc} \hline
Name & species & $t_\mathrm{grow}$ &
$t_\mathrm{disr}$ & $f_\mathrm{grow}$ &
$f_\mathrm{disr}$ & $\delta^2$ & $f_\mathrm{tot}$\\
 & & (Myr) & (Myr) & & & ($10^{-3}$) & \\ \hline
A & silicate & 30 & 3 & 0.19 & 0.45 & 2.1 & 1\\
   & graphite &  & & 0.32 & 0.24 & 3.5 & 1 \\ \hline
B & silicate & 10 & 3 & 0.065 & 0.43 & 6.8 & 1\\
   & graphite &  &  &  0.19 & 0.22 & 6.2 & 1\\ \hline
C & silicate & 30 & 10 & 0.34 & 0.16 & 4.5 & 1\\
   & graphite & & & 0.53 & 0.074 & 7.8 & 1\\ \hline
D & silicate & 10 & 10 & 0.22 & 0.14 & 20 & 1\\
   & graphite &   & & 0.33 & 0.060 & 10 & 1\\ \hline
\end{tabular}

\vspace{.1cm}
{Note:
$f_\mathrm{shock}=1-f_\mathrm{grow}-f_\mathrm{shock}$.}
\end{center}
\label{tab:ftot}
\end{table*}

\begin{table*}[t]
\renewcommand{\arraystretch}{1.2}
\vspace{-.3cm}
\caption{Models with three free parameters.}
\vspace{-.1cm}
\begin{center}
\begin{tabular}{ccccccccc} \hline
Name & species & $t_\mathrm{grow}$ &
$t_\mathrm{disr}$ & $f_\mathrm{grow}$ &
$f_\mathrm{shock}$ &
$f_\mathrm{disr}$ & $\delta^2$ & $f_\mathrm{tot}$\\
 & & (Myr) & (Myr) & & & & ($10^{-3}$) & \\ \hline
A & silicate & 30 & 3 & 0.13 & 0.62 & 0.45 & 0.98 & 1.2\\
   & graphite &  & & 0.20 & 0.76 & 0.25 & 1.8 & 1.2 \\ \hline
B & silicate & 10 & 3 & 0.038 & 0.83 & 0.47 & 2.4 & 1.3\\
   & graphite &  &  &  0.11 & 0.92 & 0.23 & 2.1 & 1.3\\ \hline
C & silicate & 30 & 10 & 0.27 & 0.80 & 0.16 & 3.0 & 1.2\\
   & graphite & & & 0.42 & 0.69 & 0.074 & 6.4 & 1.2\\ \hline
D & silicate & 10 & 10 & 0.16 & 1 & 0.15 & 10 & 1.3\\
   & graphite &   & & 0.24 & 0.99 & 0.061 & 5.8 & 1.3\\ \hline
\end{tabular}
\end{center}
\label{tab:free}
\end{table*}

\subsection{Fitting under other constraints}

In Section \ref{subsec:synthesize}, we adopted
the balance between the dust mass growth
by accretion and the dust mass loss by shock
destruction (Eq.\ \ref{eq:equil}) as a constraint.
Although this constraint is reasonable for the
dust content in the Milky Way (e.g., Inoue, 2011),
it may be useful to apply other constraints without
using Eq.\ (\ref{eq:equil}), to see
how the best-fitting parameters have been controlled
by Eq.\ (\ref{eq:equil}).

First we try to fit the MRN size distribution with
the three components under the condition that
the sum of all the fractions is unity:
$f_\mathrm{grow}+f_\mathrm{shock}+
f_\mathrm{disr}=1$. The results with this fitting are
shown in Table \ref{tab:ftot}.
The best-fitting values of $f_\mathrm{disr}$ vary
from those in Table \ref{tab:model} within
a difference of 10\% except for Model B of
silicate (25\% less). However, the
best-fitting values of
$f_\mathrm{grow}$ is broadly 1/2--2/3 of
those in Table \ref{tab:model}. As a
result, {Eq.\ (\ref{eq:equil}) is not
satisfied, and the total dust mass decreases.}

Next, we perform fitting to the MRN size
distribution with the parameters
$f_\mathrm{grow}$, $f_\mathrm{shock}$, and
$f_\mathrm{disr}$ free.
The results are shown in Table \ref{tab:free}.
Again, the values of
$f_\mathrm{disr}$ differ by only
$\lesssim 10$\% except for Model B of silicate
(18\% less). However, $f_\mathrm{grow}$ is only
$\sim 1/3$--2/3
of the values in Table \ref{tab:model}, and
$f_\mathrm{shock}$ is made large to
compensate for the decreased $f_\mathrm{grow}$.
This means that the fitting is practically dominated by
the balance between the decreased small
grains in shock destruction and the increased
small grains in disruption (shattering).
Because of the dominance of $f_\mathrm{shock}$,
{Eq.\ (\ref{eq:equil}) is not satisfied,
and} the total dust mass decreases.

\begin{figure}[t]
\includegraphics[width=0.48\textwidth]{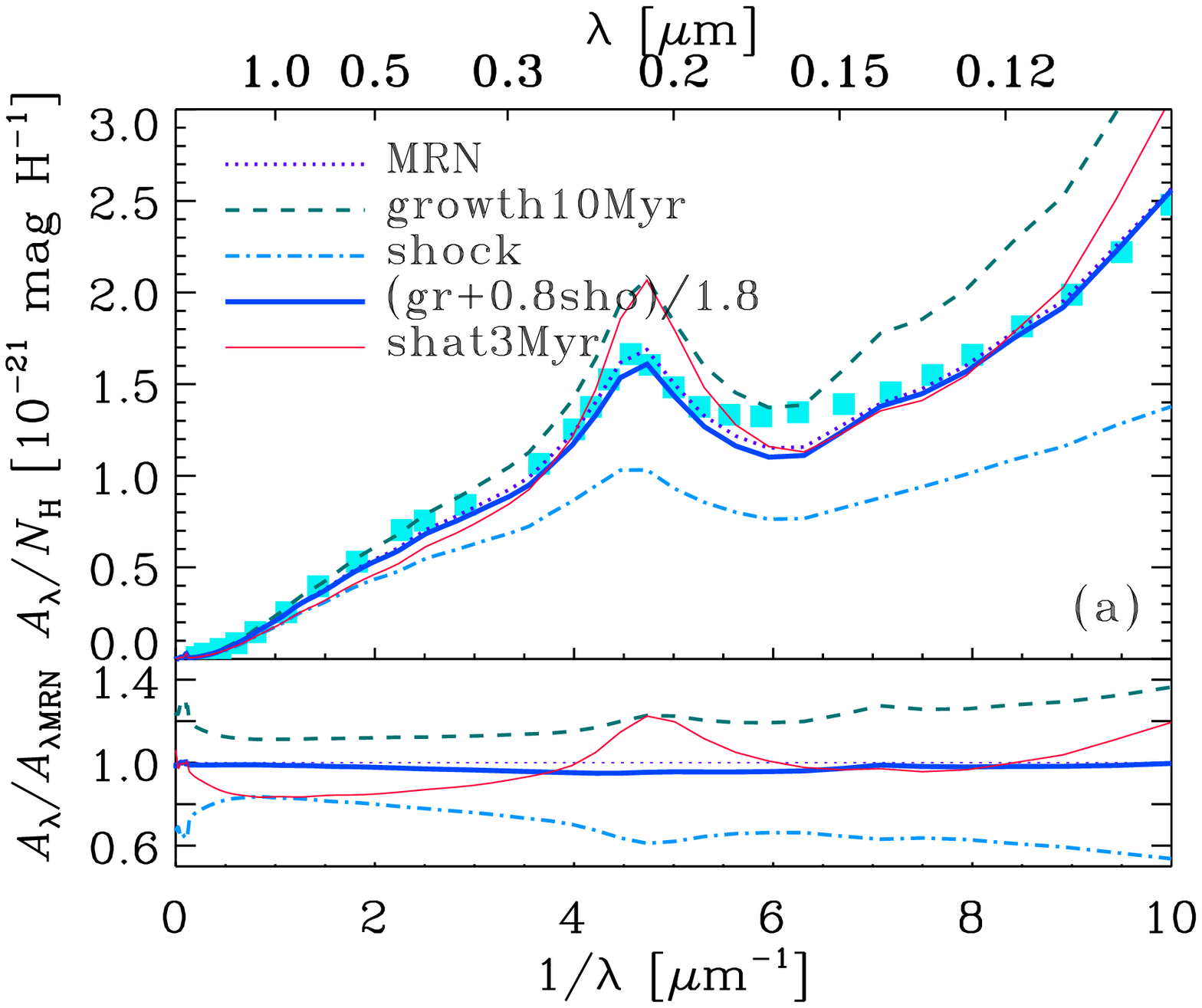}
\includegraphics[width=0.48\textwidth]{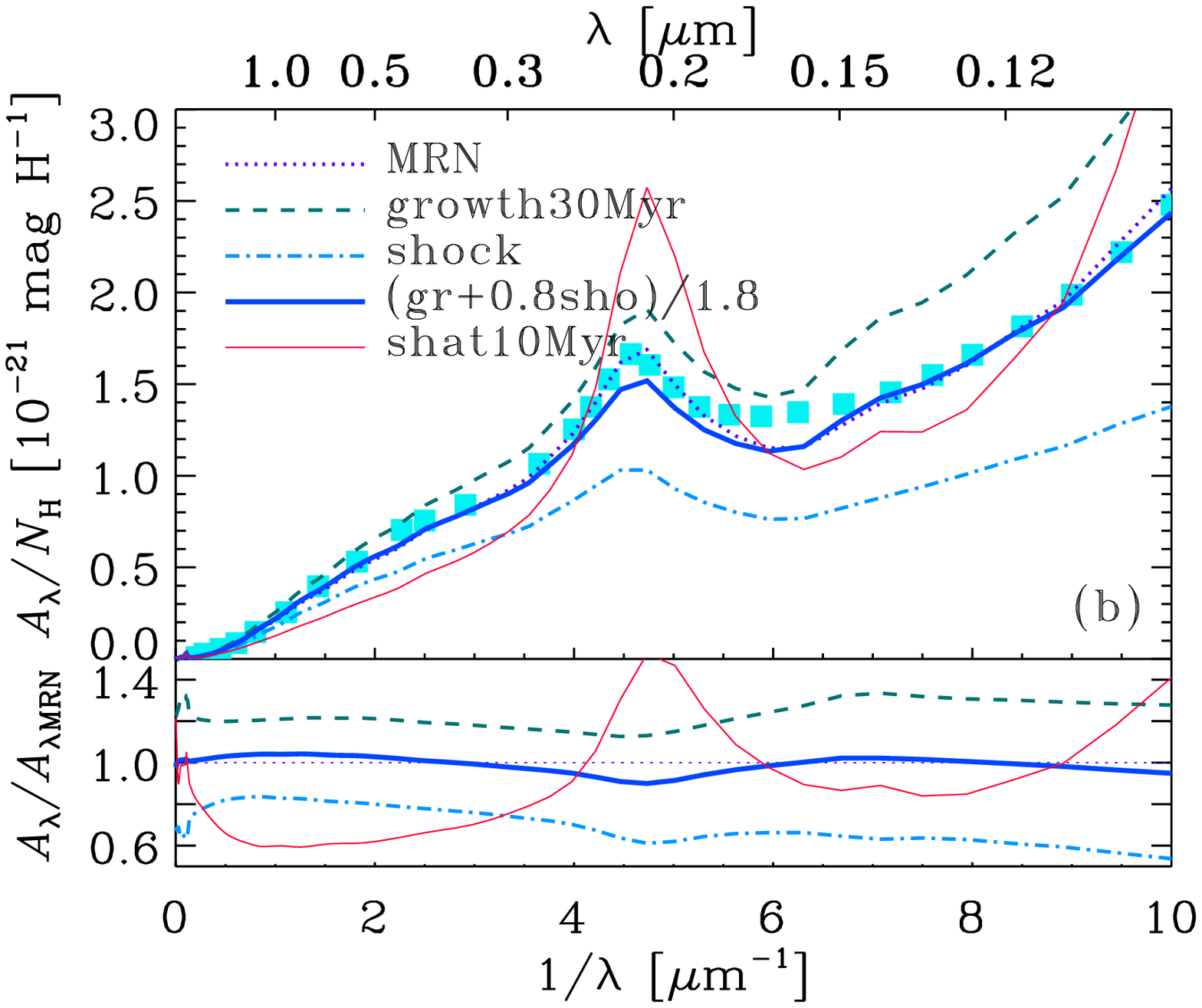}
\caption{\footnotesize
{
Upper panel: Extinction curves (extinction per
hydrogen nucleus as a function of wavelength)
calculated for
the components used for the fitting to the
grain size distribution. These components are shown
in Fig.\ \ref{fig:component}. The thin solid,
dashed, and dot-dashed lines represent
individual components processed for the following
processes: disruption (shattering) for 3 Myr in Panel (a)
and for 10 Myr in Panel (b), growth for 10 Myr in
Panel (a) and for 30 Myr in Panel (b), and shock,
respectively. The thick solid line shows the
extinction curve for
$n_\mathrm{g,s}(a)/1.8$ {(divided by 1.8 because
the component ``g,s'' contains the grain mass
1.8 times as much as the MRN)}.
The dotted line presents the extinction curve for the
MRN size distribution.
The points show the observed Milky Way extinction
curve taken from Pei (1992).
Lower panel: Ratio of extinction curves to the
extinction curve of the MRN size distribution.
The line species in the lower panel correspond to
those in the upper panel.
}
}
\label{fig:ext_component}
\end{figure}

\subsection{Extinction curve}\label{subsec:extinction}

The MRN grain size distribution is originally derived
from the Milky Way extinction curve. Therefore,
in order to check if our fitting by synthetic grain
size distributions is successful or not, it is useful
to calculate extinction curves.

Extinction curves are calculated by using the same
optical properties of silicate and graphite as those in
Hirashita and Yan (2009). The grain extinction
cross section as a function of wavelength and
grain size is derived from the Mie theory, and is
weighted for the grain
size distribution per hydrogen nucleus to obtain the
extinction curve per unit hydrogen nucleus
(denoted as $A_\lambda /N_\mathrm{H}$).
{The abundances of silicate and graphite
relative to
hydrogen nuclei are already inherent in the models
through the abundances of Si and C and $\xi$
(Section \ref{sec:model}).}

First we show the extinction curves of the
individual components, which are used to fit
the MRN size distribution, in
Fig.\ \ref{fig:ext_component}. Grain growth does
not make the extinction curve flatter in spite of
the increase of the mean grain size. The reason is
already explained in Hirashita (2012): Accretion
predominantly occurs at the smallest sizes. Since
the extinction at short wavelengths is
more sensitive to the increase of the mass of
small grains than that at long wavelengths,
the extinction curve becomes rather steeper.
Although coagulation flattens the extinction curve,
the flattening due to coagulation does not overwhelm
the steepening due to the above effect of accretion.

Shock destruction makes the extinction curve
flatter because small grains are more easily
destroyed than large grains. Grain
disruption steepens the extinction curve
because of the production of a large number
of small grains. The 0.22 $\mu$m bump
created by small graphite grains {in this model}
becomes also prominent by grain disruption. We also
show the extinction curve for the grain
size distribution
$n_\mathrm{g,s}(a)$
{(The component
$n_\mathrm{g,s}$ has a total dust mass
1.8 times as large as the initial value.
To see the difference of the extinction curve,
it would be helpful to compare under the
same dust mass, so $n_\mathrm{g,s}/1.8$ is
compared with the MRN in Fig.\ \ref{fig:ext_component}.)}

\begin{figure}[t]
\includegraphics[width=0.48\textwidth]{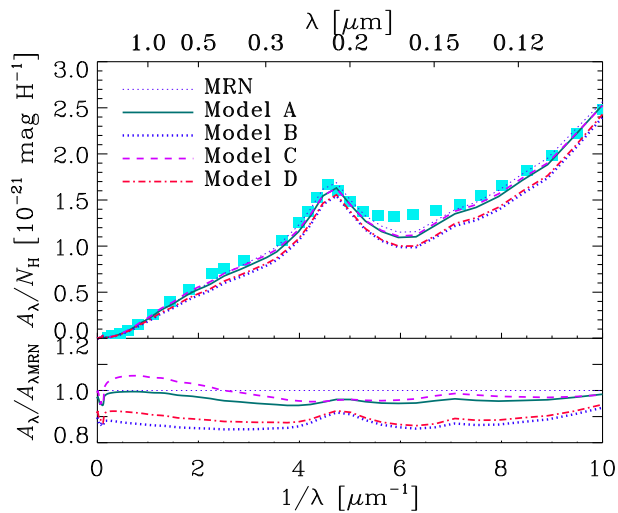}
\caption{\footnotesize
Extinction curves (extinction per hydrogen nucleus
as a function of wavelength) calculated for
Models A--D (upper
panel). The ratio to the extinction curve for the MRN
size distribution is also shown (lower panel).
The solid, thick dotted, dashed, and
dot-dashed lines represent Models A, B, C, and D,
respectively. The thin dotted line shows the
extinction curve for the MRN size distribution.
The points show the observed Milky Way extinction
curve taken from Pei (1992).
}
\label{fig:extinc}
\end{figure}

In Fig.\ \ref{fig:extinc}, we show the extinction curves
calculated for Models A--D. First of all, we confirm
that the MRN size distribution reproduces the
observed extinction curve (some small deviations
can be fitted further if we adopt a more detailed
functional form of the grain size distribution, which is
beyond the scope of this paper; see
Weingartner and Draine, 2001, for a detailed
fitting). Comparing the
extinction curve for the MRN size distribution and
those for Models A--D, we observe that the
extinction curve is reproduced within a
difference of $\sim 10$\%. Models A and C
are successful, while Models B and D systematically
underproduce the MRN extinction curve (although
the difference is small).
{The underprediction by
$\sim 10$\% in
Models B and D occurs because
$f_\mathrm{tot}$ is $\sim 0.9$.}

\begin{figure}[t]
\includegraphics[width=0.48\textwidth]{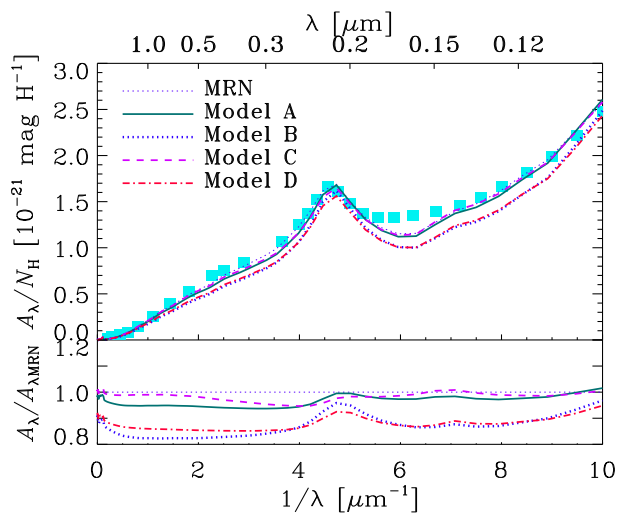}
\caption{\footnotesize
Same as Fig.\ \ref{fig:extinc} but we
use the mean values between silicate and graphite
for $f_\mathrm{grow}$ and $f_\mathrm{disr}$.
}
\label{fig:extinc_mean}
\end{figure}

In the above, we adopted different values for
$f_\mathrm{grow}$ and $f_\mathrm{disr}$ between
silicate and graphite. As mentioned in
Section~\ref{subsec:synthesize}, if both species
are well mixed in the ISM, they would have
common values of these parameters.
In Fig.\ \ref{fig:extinc_mean}, we show the
extinction curve by taking the average of
the values for silicate and graphite
(for example, $f_\mathrm{grow}=0.35$
and $f_\mathrm{disr}=0.35$
for Model A). We find that the difference
between Figs.\ \ref{fig:extinc} and \ref{fig:extinc_mean}
is small. Therefore, the mean values work to reproduce
the Milky Way extinction curves. The mean values
are in the range of $f_\mathrm{grow}=0.3$--0.5 and
$f_\mathrm{disr}=0.1$--0.4.
We conclude that the synthetic grain size distributions
with these parameter ranges reproduce the Milky Way
extinction curve.

\section{Conclusion}\label{sec:conclusion}

In our previous papers (Nozawa \textit{et al.}, 2006;
Hirashita and Yan, 2009; Hirashita, 2012), we showed
that dust grains are quickly processed by shock
destruction, disruption, and grain growth. In this
paper, thus, we have examined if the MRN grain size
distribution, which is believed to
represent the grain size distribution in the Milky Way,
can be reproduced by the processed grain
size distributions. We ``synthesized'' the grain size
distribution by summing the processed grain size
distributions under the condition that  the decrease of
dust mass by shock destruction is compensated by
grain growth. We have found that the synthetic
grain size distribution can reproduce the MRN grain
size distribution in the sense that the deficiency of
small grains by grain growth and shock destruction
can be compensated by the production of small grains
by disruption.
The values of the fitting parameters
indicate that, among the processed grains,
30--50\% is growing in dense medium,
20--40\% is being destroyed by shocks in diffuse medium,
and 10--40\% is being shattered in diffuse medium
(the percentage shows the relative importance of
each process).
The extinction curves calculated
by the synthesized grain size distributions reproduce
the observed Milky Way extinction curve within a difference
of $\sim 10$\%. This means that our idea of
synthesizing the grain size distribution based on
major processing mechanisms (i.e.,
grain growth, shock destruction, and disruption)
is promising
as a general method to ``reconstruct'' the
extinction curve.

\acknowledgments{
We are grateful to C. Wickramasinghe and an anonymous
reviewer for helpful comments in their review of this paper.
HH has been
supported through NSC grant 99-2112-M-001-006-MY3.
TN has been supported by World Premier
International Research Center Initiative (WPI Initiative), MEXT,
Japan, and by the Grant-in-Aid for Scientific Research of the
Japan Society for the Promotion of Science (20340038,
22684004, 23224004). }


\email{H. Hirashita (e-mail: hirashita@asiaa.sinica.edu.tw)
and T. Nozawa}

\label{finalpage}
\lastpagesettings

\end{document}